\documentclass{PoS}
\usepackage{amsmath,amssymb,graphicx}
\input{definitions.command}

\usepackage{floatrow}
\newfloatcommand{capbtabbox}{table}[][\FBwidth]

\title{Hosotani mechanism on the lattice}

\ShortTitle{Hosotani mechanism on the lattice}

\author{\speaker{Guido Cossu}\\
  Theory Center, IPNS, High Energy Accelerator Research Organization (KEK),\\
  Tsukuba, Ibaraki 305-0810, Japan\\
  E-mail: \email{cossu@post.kek.jp}}

\author{Etsuko Itou\\
  Theory Center, IPNS, High Energy Accelerator Research Organization (KEK),\\
  Tsukuba, Ibaraki 305-0810, Japan\\
  E-mail: \email{eitou@post.kek.jp}}

\author{Hisaki Hatanaka\\
  School of Physics, KIAS,\\ Seoul 130-722, Republic of Korea\\
  E-mail: \email{hatanaka@kias.re.kr}}

\author{Yutaka Hosotani\\
  Department of Physics, Osaka University,\\ Toyonaka, Osaka 560-0043, Japan\\
  E-mail: \email{hosotani@phys.sci.osaka-u.ac.jp}}

\author{Jun-Ichi Noaki\\
  Theory Center, IPNS, High Energy Accelerator Research Organization (KEK),\\
  Tsukuba, Ibaraki 305-0810, Japan\\
  E-mail: \email{noaki@post.kek.jp}}

\abstract{We explore the phase structure and symmetry breaking in four-dimensional SU(3) gauge theory with one spatial compact dimension on the lattice in the presence of fermions in the adjoint and fundamental representations with general boundary conditions. The eigenvalue phases of Polyakov loops and the associated susceptibility are measured on $16^3 \times 4$ lattice. We establish a correspondence between the phases found on the lattice and the gauge symmetry breaking by the Hosotani mechanism.}

\FullConference{31st International Symposium on Lattice Field Theory - LATTICE 2013\\
  July 29 - August 3, 2013\\
  Mainz, Germany}

\begin{document}

\section{Introduction}

Symmetry breaking mechanisms play a central part in the 
unification of gauge forces. The gauge symmetry of a unified theory must
be partially and spontaneously broken at low energies to describe the nature.
In the standard model (SM) of electroweak interactions, the Higgs scalar field
induces the symmetry breaking.

Among the several mechanism for gauge symmetry breaking there is the intriguing scenario of
dynamical breaking by adding compact extra dimensions.  
In brief, when the extra dimensional space is not simply-connected, the
non-vanishing phases $\theta_H$ of the Wilson line integral of gauge
fields along a non-contractible loop in these extra dimensions 
can break the symmetry of the vacuum at one loop level~\cite{YH1,Davies1,Davies2,YH2}.
These phases $\theta_H$ are the 
Aharonov-Bohm (AB) phases in the extra dimensional space, which, despite its
vanishing field strengths, affect physics leading to gauge symmetry breaking.
This is the so called the Hosotani mechanism where  the 4D Higgs boson is a part of gauge fields in higher dimensions.  
The values of $\theta_H$ are determined dynamically.
Recently, the Hosotani mechanism has been applied to the electroweak interactions 
\cite{Nomura1,Csaki2,ACP,Csaki3,MSW, HOOS,HNU}.

It should be pointed out that the Hosotani mechanism as a
mechanism of gauge symmetry
breaking has been so far established only in perturbation theory. 
It is based on the evaluation of
the effective potential $V_\eff (\theta_H)$ at the one-loop level.  
It is still not clear whether the mechanism operates at the non-perturbative level. 
This work is a first investigation on the non-perturbative realization
of the Hosotani mechanism using lattice calculations.
We take advantage of the fact that the Hosotani mechanism works
in any dimensions such as $R^n \times S^1$, so we focus on the
four-dimensional case ($R^3 \times S^1$) in which the lattice gauge theory has been firmly established. 

In this work, we would like to point out the connection between the
phases identified by Cossu and D'Elia ~\cite{Cossu} (in a work inspired by the
semi-classical study~\cite{Unsal}) and the Hosotani mechanism~\cite{HosotaniGUT2012}.
We also refine the connection by generalizing the boundary conditions for 
fermions in the fundamental representation. 
The rest of the report is presenting the theoretical background 
in Sect.~\ref{sec:ABphasesInContinuum} and the lattice calculations 
in Sect.~\ref{sec:LatticeResults}. This proceeding is a summarized version of the 
full paper recently published online~\cite{Cossu:2013ora}.

\section{Continuum gauge theory on $R^{d-1} \times S^1$}
\label{sec:ABphasesInContinuum}

As the simplest realization of the Hosotani mechanism, we consider SU(3) gauge theory coupled with 
fermions in the fundamental representation ($\psi_\fund$)
and/or in the adjoint representation ($\psi_\ad$) in $d$-dimensional flat space-time with one spatial 
dimension compactified on $S^1$ \cite{Hatanaka1999,Hosotani2005}.   
The circle $S^1$ has coordinate $y$ with a radius $R$ so that
$y \sim y + 2\pi R$.  In terms of these quantities the Lagrangian density is given by:
\beeq
\cL = - \frac{1}{2} \Tr F_{MN} F^{MN} + \bar \psi_\fund (\cD_\fund - m_\fund) \psi_\fund
+ \Tr \bar \psi_\ad (\cD_\ad - m_\ad) \psi_\ad 
\label{Lagrangian1}
\eneq
where $\cD_\fund$ and $\cD_\ad$ denote covariant Dirac operators.
The gauge potentials $A_M = (A_\mu, A_y)$ ($\mu = 1, \cdots, d-1$) and
fermions $\psi_\fund, \psi_\ad$ satisfy the following boundary conditions:
\begin{align}
\begin{split}
A_M ( x, y + 2\pi R) &= V A_M ( x, y)  V^{-1},\\
\psi_\fund  ( x, y + 2\pi R) = e^{i \alpha_\fund} \, V \, \psi_\fund  ( x, y),\quad 
&\psi_\ad ( x, y + 2\pi R)  = e^{i \alpha_\ad} \, V \, \psi_\ad  ( x, y)  V^{-1}, 
\end{split}
\label{eq:BoundCond_1}
\end{align}
where $V \in SU(3)$.  
With these boundary conditions the Lagrangian density
is single-valued on $S^1$, namely $\cL (x, y + 2\pi R) = \cL (x, y )$, so that physics is 
well-defined on the manifold $R^{d-1} \times S^1$. It has been proven (see \cite{YH2}) 
that physics is independent of $V$ at the quantum level so we adopt $V=I$ hereafter. 

There is a residual gauge invariance given the boundary conditions (\ref{eq:BoundCond_1}).
Under a gauge transformation $\Omega$, 
the boundary condition (\ref{eq:BoundCond_1}) with $V=I$ is maintained, provided $\Omega(x, y+ 2\pi R) = \Omega(x, y)$. 
The eigenvalues of $W= P \exp \left( i g \int_0^{2\pi R} dy \, A_y(x,y) \right)$ are gauge invariant.  They are denoted as
$\big\{ e^{i \theta_1},  e^{i \theta_2},   e^{i \theta_3} \big\}$ where $\sum_{j=1}^3 \theta_j = 0 ~ (\bmod ~ 2\pi)~$.
Constant configurations of $A_y \not= 0$ with $A_\mu=0$ yield vanishing
field strengths $\langle F_{MN}\rangle=0$, but in general give 
$W \neq I$, or nontrivial $\theta_H$.
We stress that this class of configurations is not gauge equivalent to $A_M =0$ if we want to keep the boundary conditions constant.
The $\theta_j $'s are the elements of AB phase $\theta_H$ in the 
extra dimension. These are the dynamical degrees of freedom 
of the gauge fields affecting physical quantities
as in the Aharonov-Bohm effect in quantum mechanics.

\subsection{Symmetry breaking\label{sec:sym-breaking}}

To see the effect of the AB phases on the spectrum of gauge bosons we expand the fields 
of the SU(3) gauge theory on $R^{d-1} \times S^1$ in Kaluza-Klein modes of the extra-dimension:
\cite{YH2} where each KK mode has the following mass-squared in the $(d-1)$-dimensional 
space-time.
\begin{align*}
\begin{split}
A_\mu^{(n)} :& \left(m_A^{(n)}\right)_{jk}^2 = \frac{1}{R^2} \Big( n + \frac{\theta_j - \theta_k}{2\pi} \Big)^2 ~,\\
\psi_\fund^{(n)} : \left(m_\fund^{(n)}\right)_j^2 = \frac{1}{R^2} \Big( n +  \frac{\theta_j + \alpha_\fund}{2\pi} \Big)^2 
   +& m_\fund^2 ~, 
\psi_\ad^{(n)} : \left(m_\ad^{(n)}\right)_{jk}^2=\frac{1}{R^2} \Big( n +  \frac{\theta_j  - \theta_k 
     + \alpha_\ad}{2\pi} \Big)^2     + m_\ad^2 ~.
\label{spectrum1}
\end{split}
\end{align*}

\begin{figure}[h]
  \centering
  \includegraphics[width=0.9\textwidth]{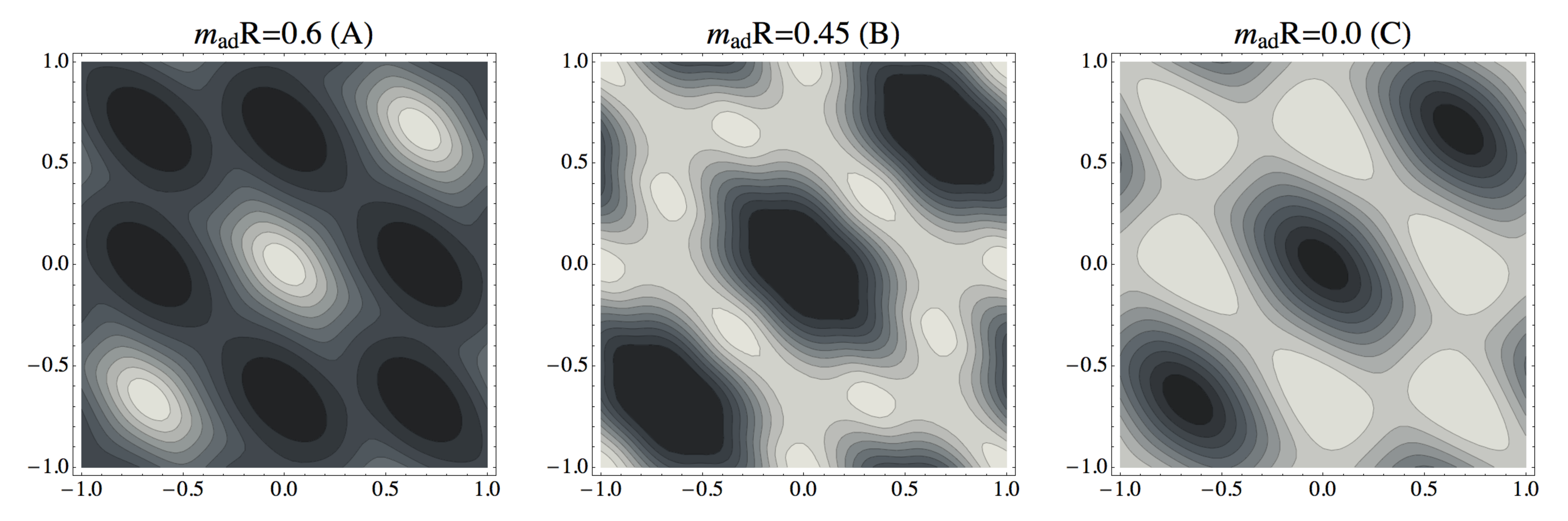}
  \caption{Effective potential for the case of $N_\ad = 2$ adjoint fermion 
    with periodic boundary condition ($\alpha_{\rm ad}=0$) for 
    the values of $m_{\rm ad} R$ in $d=4$. They are corresponding to 
    the $X$ phase, the $B$ phase and the $C$ phase, respectively.
    Lower values of $V_{\eff}$ are indicated by lighter colors.}
  \label{fig:ad-mass}
\end{figure}

In particular, from the gauge boson mass of the zero-mode $(m_A^{(0)})^2$,
we can discuss the remaining gauge symmetry realization after the compactification.
Because the mass is given by the difference $\theta_j-\theta_k$, it is
classically expected that the mass spectrum becomes SU(3) asymmetric
unless $\theta_1=\theta_2=\theta_3\ ({\rm mod}\ 2\pi ) $.
However, as a dynamical degree of freedom, $\theta_H$ has quantum fluctuation. 
In the confined phase, these fluctuations are large enough for the SU(3)
symmetry to remain intact.
For a moderate gauge coupling and sufficiently small $R$, $\theta_H$ would take 
nontrivial values to break SU(3) symmetry depending on the fermion content.
To determine which value of $\theta_H$ is realized at the quantum level, 
it is convenient to evaluate the effective potential $V_\eff (\theta_H)$,
whose global minimum is given by the vacuum expectation values (VEVs) of $\theta_H$.
We show the plots for the two flavors of adjoint fermions case in figure~\ref{fig:ad-mass} in order to 
compare with the lattice simulations.
In lattice simulations one measures the VEVs of 
$P_3$ and $P_8$. The absolute value of $ P_3 $ is strongly affected by
quantum fluctuations of $\theta_H$ and is reduced at strong gauge couplings.
The phase of $ P_3$, on the other hand, is less affected by quantum fluctuations in the weak 
coupling regime so that transitions from one phase to another should be
seen as changes in the 
phase of $ P_3 $.  Indeed, this is precisely what has been found in ref.~\cite{Cossu}.
The classifications of the phases are summarized in Table~\ref{table-phase}, where
we also include the confined phase, denoted by $X$, in which $\theta_H$ fluctuate and
take all possible values. 

\begin{table}[bth]

\caption{Classification of the location of the global minima of $V_\eff (\theta_H)$.
In the last column the names of the corresponding phases termed in
 ref.~\cite{Cossu} are also listed for $X, A, B$ and $C$.}
\begin{center}
\renewcommand{\arraystretch}{1.5}
\begin{tabular}{|c|c|c|c|c|}
\hline
 & $\theta_H=(\theta_1, \theta_2, \theta_3)$ with permutations
 & $P_3$ & $P_8$ & Global Symmetry, Phase\\
\hline
$X$ & Large quantum fluctuations & $0$ &
	     $-\frac{1}{8}$ & SU(3), confined \\
\hline
$A_1; A_{2,3}$ & (0,0,0); $(\pm \twothird \pi, \pm \twothird \pi, \pm \twothird \pi)$ 
& 1; $e^{\pm 2\pi i /3}$ & 1 & SU(3), deconfined\\
\hline
$B_1$;$B_{2,3}$ & $(0,\pi, \pi)$; $(\pm \twothird \pi, \mp \onethird \pi, \mp \onethird \pi)$ 
& $- \onethird$; $\onethird e^{\mp \pi i/3}$ & 0 & SU(2) $\times$ U(1), split \\
\hline
$C$  & $(0,\twothird \pi , - \twothird \pi)$ & 0
& $- \frac{1}{8}$ & U(1) $\times$ U(1), reconfined \\
\hline
\end{tabular}
\end{center}
\label{table-phase}
\end{table}

\section{Lattice results\label{sec:LatticeResults}}

We compute Polyakov loops $P_3$ and $P_8$
on the $16^3\times 4$ volume gauge configurations
sampled with the weight 
$e^{-S_g - S_f}$. 
The transition points were determined using the susceptibility $\chi_\Omega = N_x^3 \left( \langle \Omega^2 \rangle - \langle \Omega
\rangle^2 \right)$ of the observable $\Omega\in \{|P_3|,P_8\}$ which
scales with the lattice volume at the phase transitions.
 In connection to the perturbative results, where the relevant parameter 
 is $m_\fund R$ or $m_\ad R$, increasing $\beta$ has the effect of 
decreasing those parameters, due to the running of the renormalized 
fermion mass in the lattice unit.
We estimate statistical errors by employing the jackknife method 
with appropriate bin sizes to incorporate any auto-correlations.

\subsection{Phase structure with adjoint fermions}\label{AdjointSection}

In the numerical simulation for $(N_{\rm ad},N_\fund) = (2,0)$, 
we use bare masses $ m_\ad a = ma =0.05$ and $0.10$ 
changing $\beta$ covering the range $5.3 \le \beta \le 6.5$.
Periodic boundary condition is used ($\alpha_\ad=0$) 
in the compact direction, which is different 
from the case with anti-periodic boundary conditions (finite temperature) 
where only the confined and deconfined phases are 
realized~\cite{Karsch:1998qj}.
The essentially same setup is included in the study of ref.~\cite{Cossu}.
To explore the phase structure in heavier mass region, we also examine 
bare masses $ m_\ad a = ma =0.50$ and $0.80$ 
for the range of $5.5 \le \beta \le 9.8$ and $5.5 \le \beta \le 20.0$,
respectively. 

\begin{figure}[h]
  \centering
  \includegraphics[clip=true,width=0.5\textwidth]{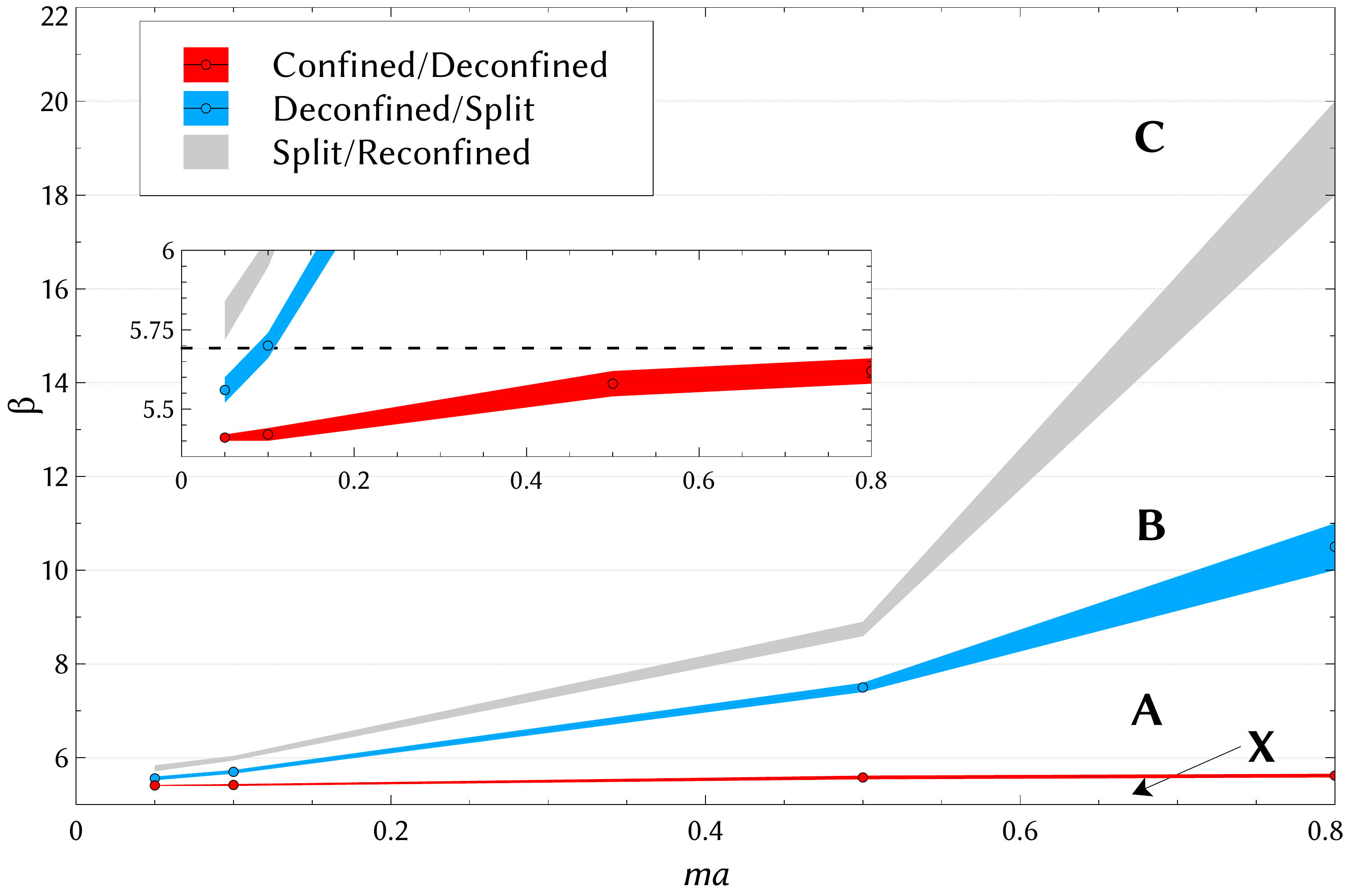}
   {
      \caption{Phase diagram for the $N_\ad =2$ adjoint fermion system with periodic 
        boundary condition in the compact dimension. 
        In the window, the $X$-$A$ transition line is compared with the 
        pure gauge case (dashed line)~\cite{Fukugita1989}. }
    }
    \label{fig:phase-diagram}
\end{figure}

We summarize the phase space and the phase transition points obtained in figure~\ref{fig:phase-diagram}.
For further discussion on the properties of these transitions, 
a more detailed study on the finite size scaling has to be done. 
 
Because the $B$-$C$ transitions are hard to observe clearly from the Polyakov
loops or the susceptibilities, we estimate empirically the interval where the transition occurs by inspection of the $\theta$s distributions. Due to the subjective character of the analysis, we do not quote any error, just an interval where the transition is occurring.

\subsection{Phase structure with fundamental fermions}
\label{FundamentalSection}

\begin{figure}[ht]
 \centering
  \includegraphics[width=0.35\textwidth,clip]{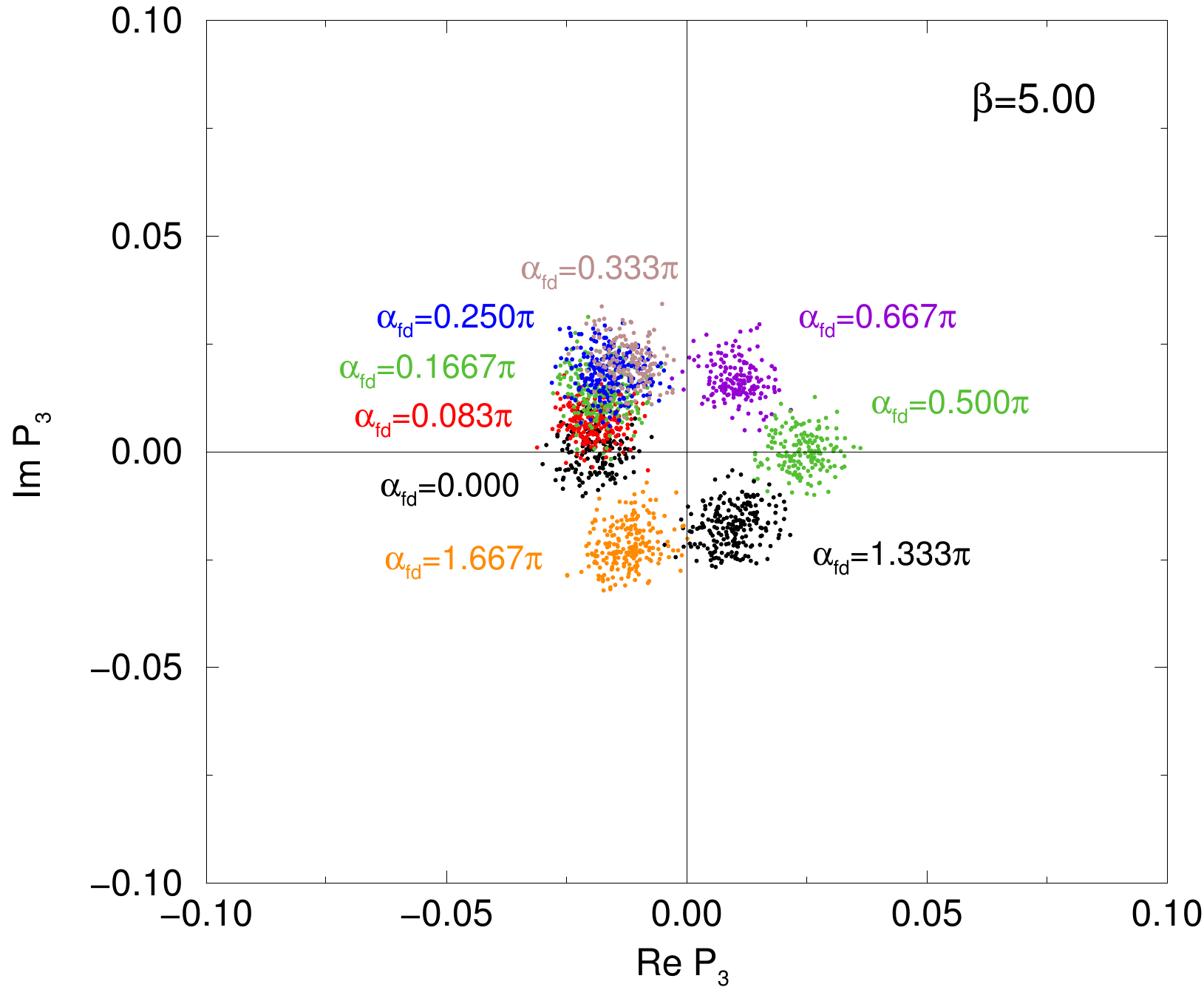}
  \includegraphics[width=0.31\textwidth,clip]{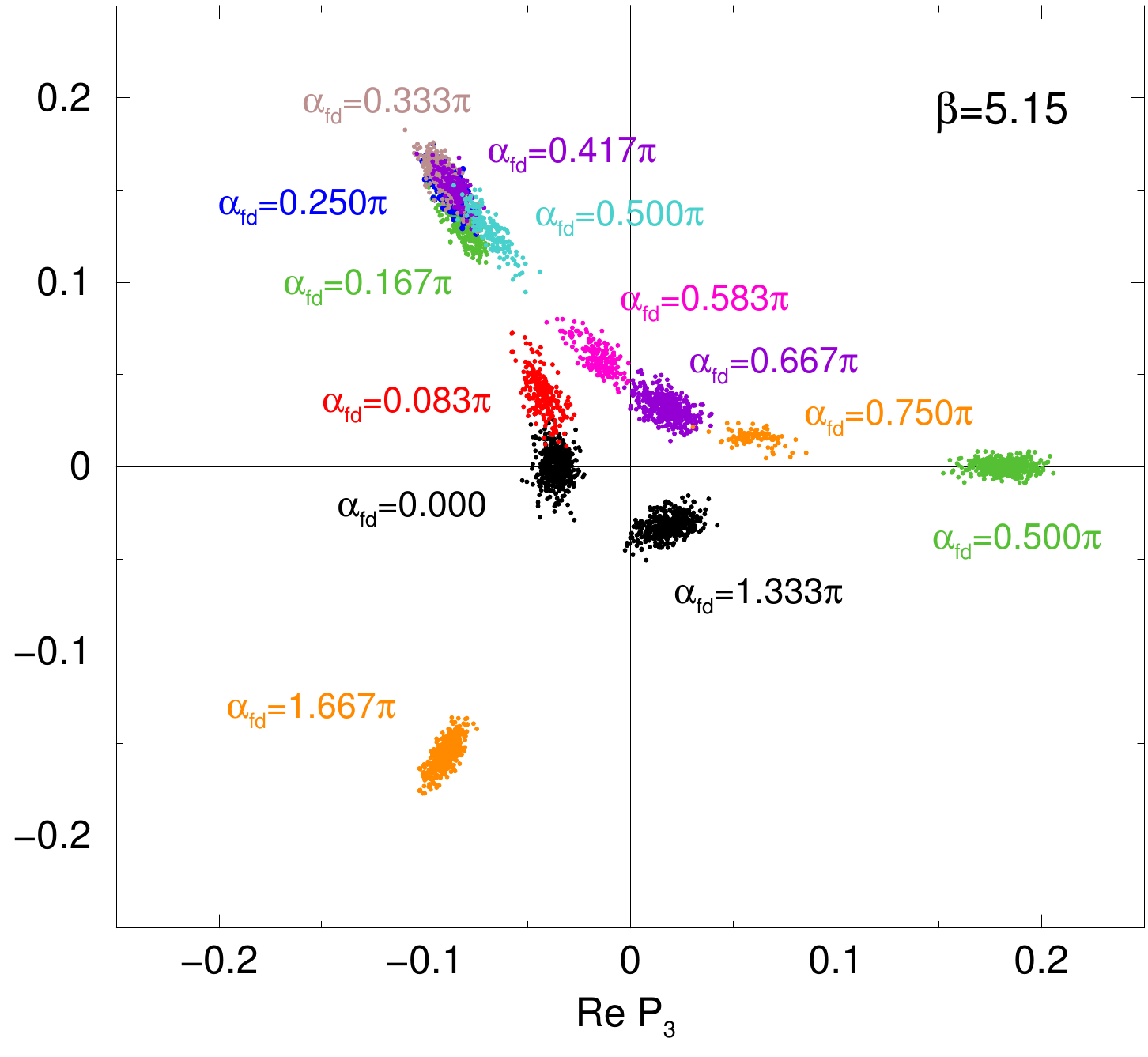}
  \includegraphics[width=0.31\textwidth,clip]{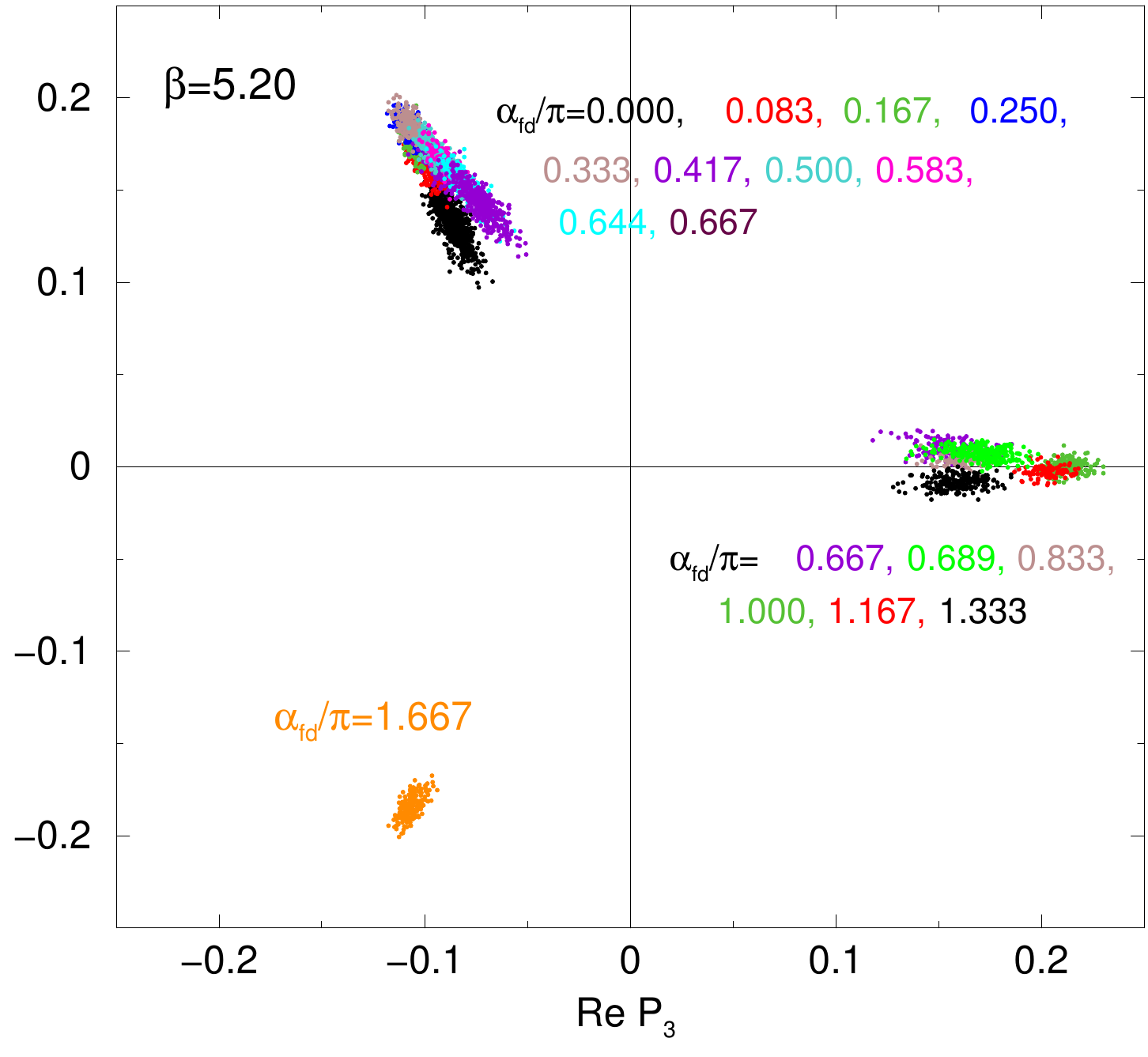}
  \caption{(First three panels) Distributions of $P_3$ obtained on gauge ensembles
    with a variation of $\alpha_\fund$ for $\beta=5.00$ (left), 
    $5.15$ (center) and $5.20$ (right). The degrees of $\alpha_\fund$ used in
    the calculation are indicated with the corresponding data. 
    Overall, data points with same degrees of $\alpha_\fund$ are 
    indicated by same colors.}
  \label{fdm_deconfined}
\end{figure}

We study the dependence of $P_3$ and $P_8$ on the boundary phase $\alpha_\fund$ 
for several values of $\beta$ in the presence of fundamental fermions with U(1) phase
 $\alpha_\fund$ as the boundary condition.
This setup is formally equivalent to finite
temperature QCD with an imaginary chemical potential $\nu= \pi
+\alpha_\fund$.
To test the perturbative predictions we 
carry out a numerical simulation with $(N_{\rm ad},N_\fund) = (0,4)$.
Since we are interested in the symmetries of the Polyakov loop,
we determine the locations of the transition points 
by the peak points of $\chi_{|P_3|}$, contrary to the previous works where the 
chiral condensate was used to locate the critical points. The resulting distributions of $P_3$ are shown in
Fig.~\ref{fdm_deconfined}.

Having confirmed the Roberge-Weiss periodic structure \cite{Roberge1986} and using the symmetry 
about $\alpha_\fund=\pi/3$, 
we concentrate on the region $0\le\alpha_\fund< \pi/3$
to determine the $A$-$B$ (or, confined-deconfined) transition points. 
For $\alpha_\fund=n\pi/12$ with $n=0,1,2,3,4$, we investigate the 
susceptibilities of $|P_3|$ and $P_8$ along with the analysis in the 
previous section. We obtain the well known Roberge-Weiss phase diagram 
in terms of the phases predicted by the Hosotani mechanism. The deconfined phases
at high $\beta$ are identified as the $A_i$ phases of table~\ref{table-phase}.

\subsection{Eigenvalues of the Wilson line \label{sec:Directmeasure}}

In the previous sections we identified four phases in the case of adjoint fermions with periodic boundary 
condition.
The comparison of the measured values for $P_3$ and $P_8$ with the ones 
listed in Table~\ref{table-phase} suggests that these phases are related to the Hosotani mechanism.
In order to clarify the connection of these phases with the perturbative 
effective potential predictions we show 
the results of the eigenvalues of the Wilson line wrapping around the compact dimension.

In this analysis care must be taken in order to disentangle the effect of the Haar measure for 
SU(3)~\cite{Bruckmann:2010tz} $\prod_{i>j} \sin^2 \frac{\theta_i - \theta_j}{2}$.
This measure term gives a strong repulsive force for the eigenvalues.
We estimated numerically the effect of this term and renormalized the lattice results accordingly.
The $A$, $B$ and $C$ phases should show  different degeneracy of eigenvalues as shown in table~\ref{table-phase}.

\begin{figure}[th]
 \centering
  \includegraphics[clip=true, width=0.24\columnwidth]{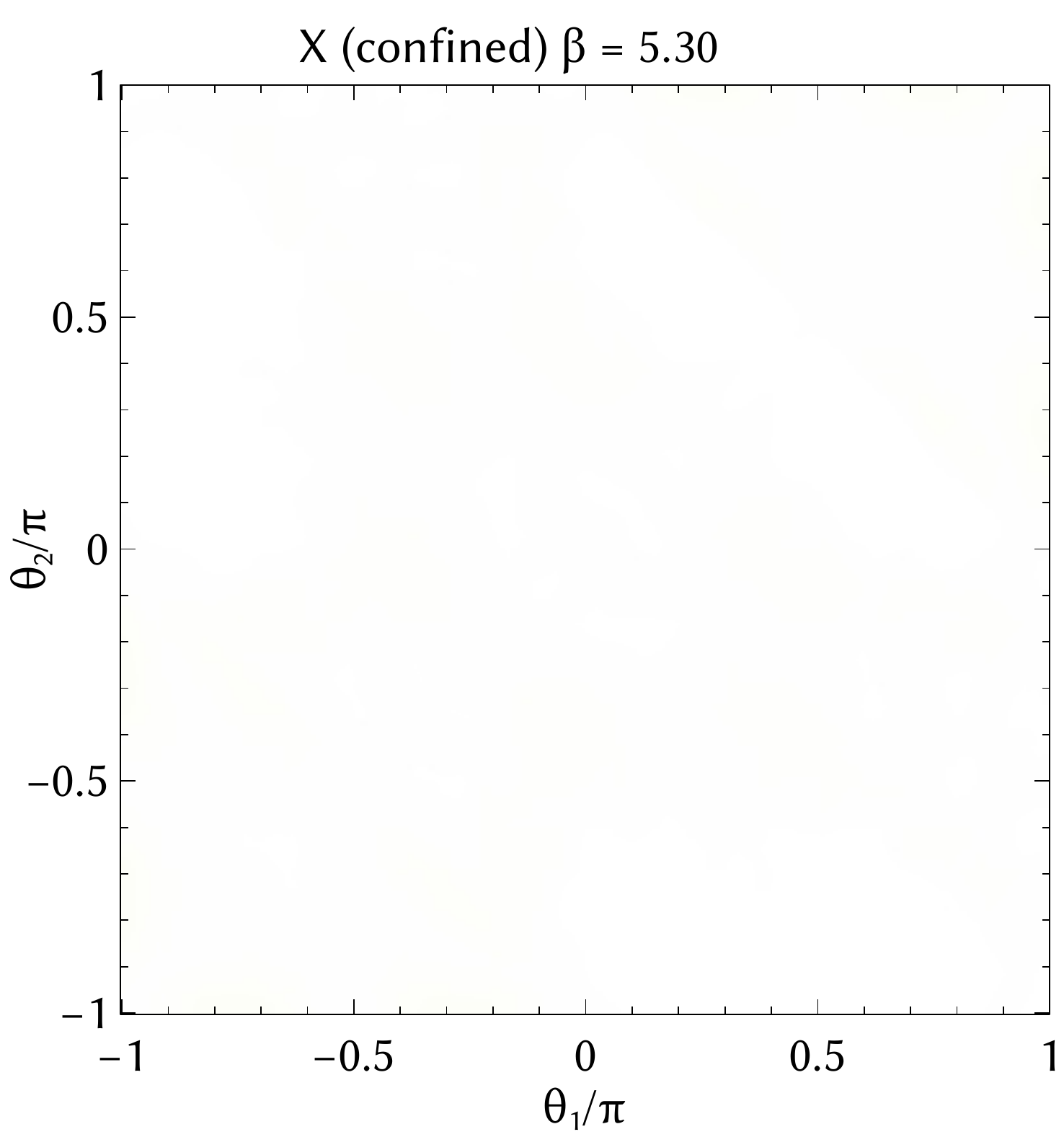}
  \includegraphics[clip=true, width=0.24\columnwidth]{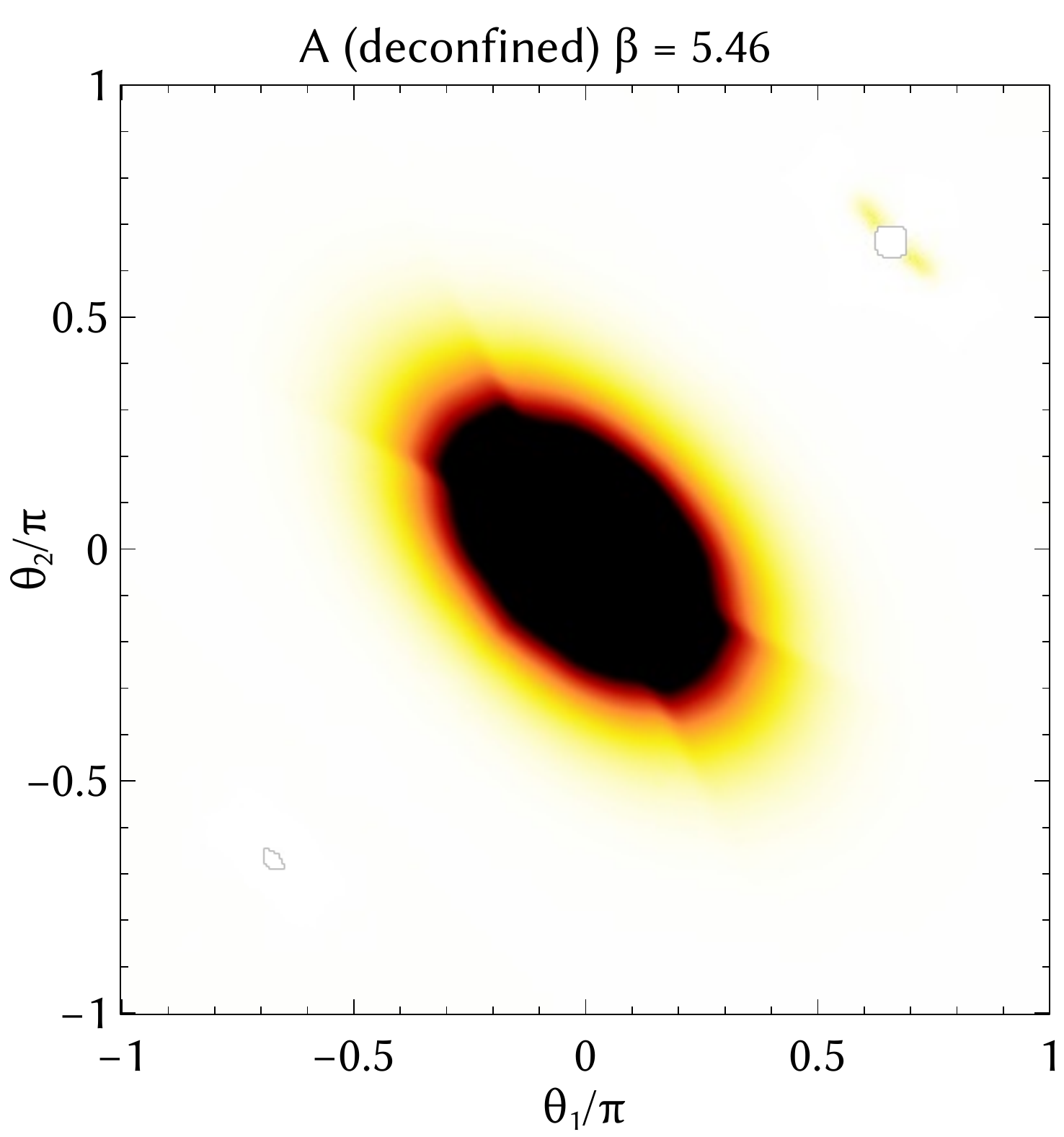}
  \includegraphics[clip=true, width=0.24\columnwidth]{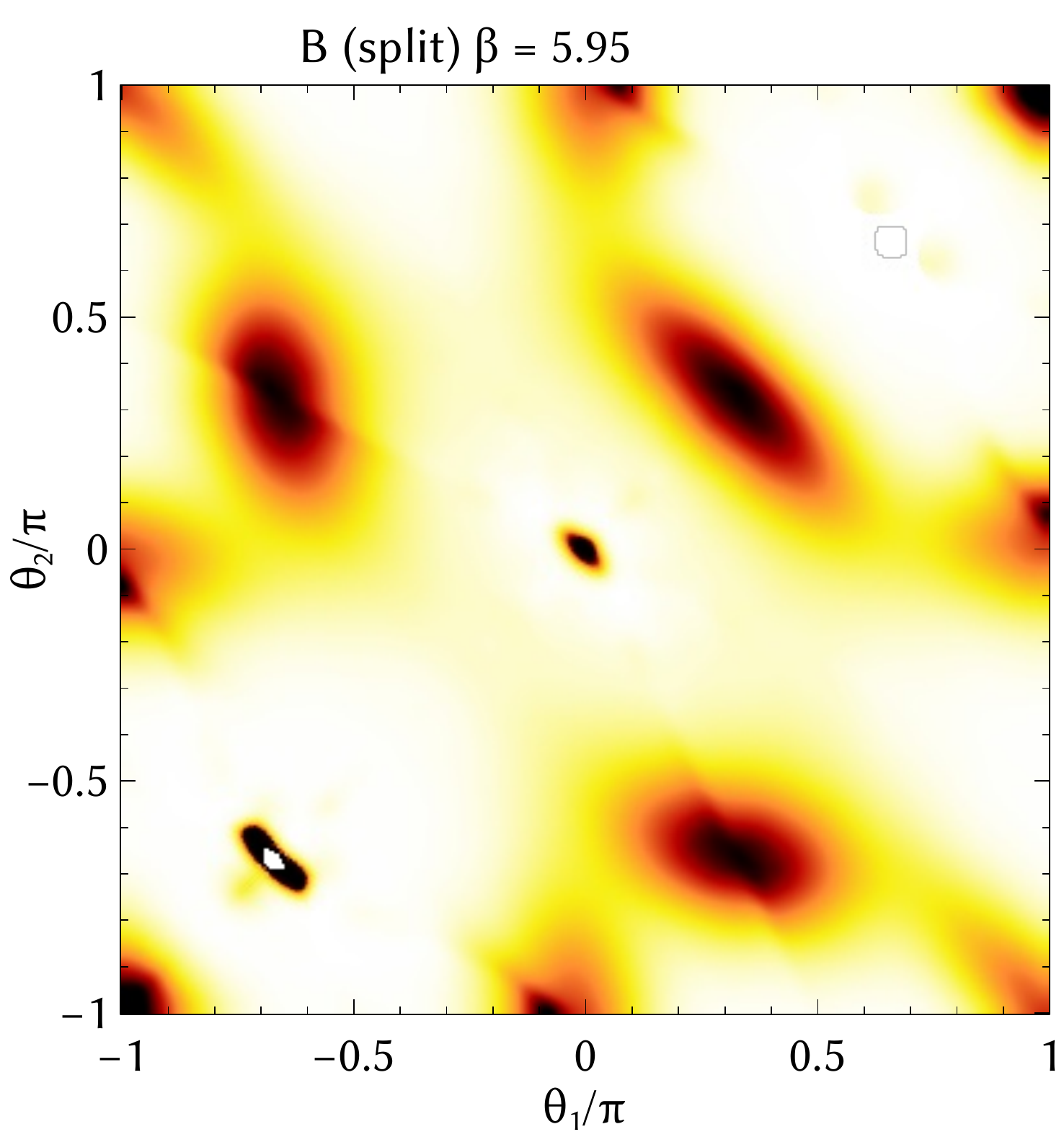}
  \includegraphics[clip=true, width=0.24\columnwidth]{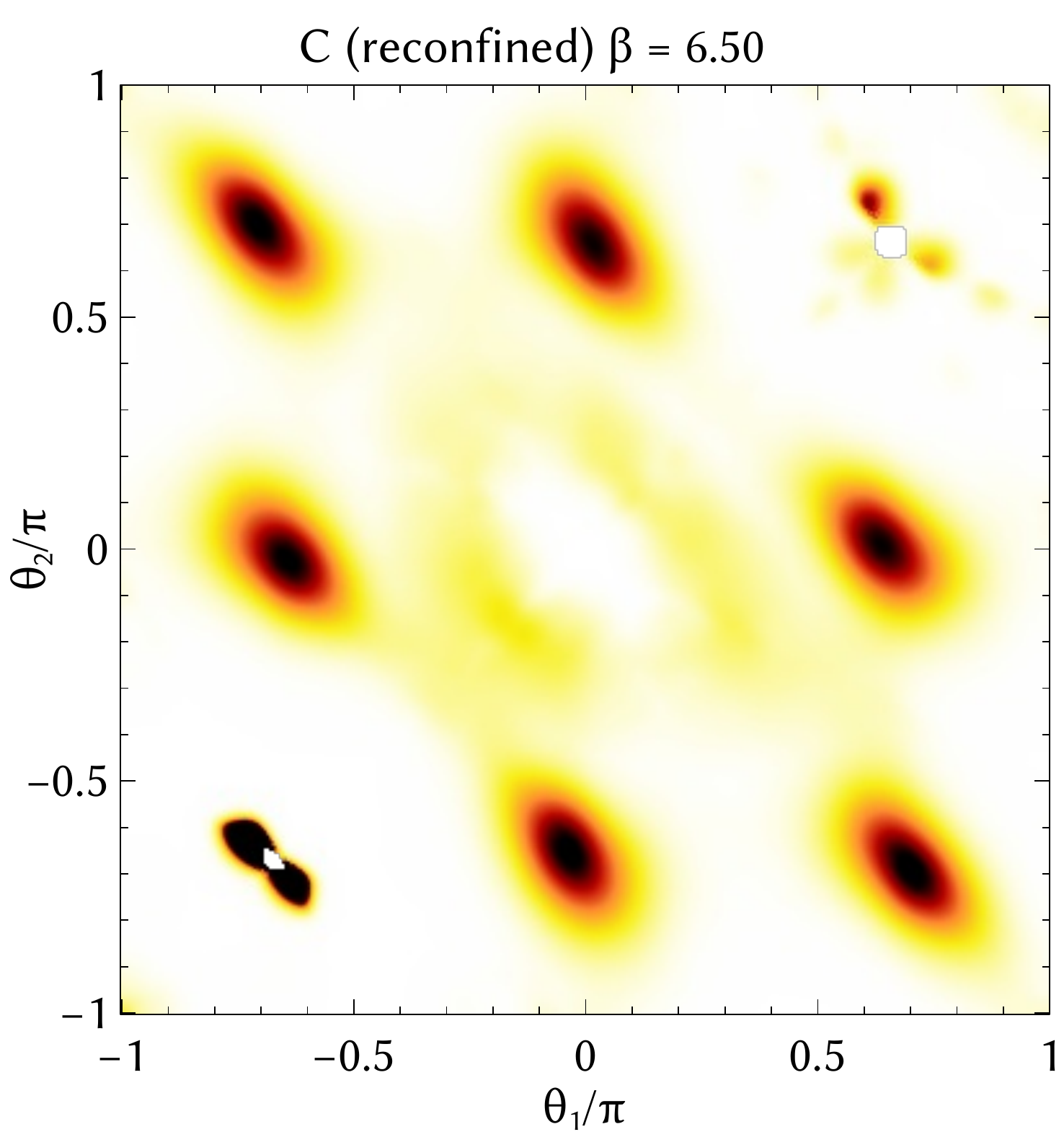}
 \caption{Density plots at several $\beta$'s for the Polyakov Loop
  eigenvalues (in the $\theta_1/\pi$ - $\theta_2/\pi$ plane).
  Here the original data is divided by the Haar measure distribution. 
  From left to right, the panels correspond to the $X$, $A$, $B$ and 
  $C$ phases.
  The first panel is white as a result of the calculation.
  Darker colors denote the highest density regions.}
 \label{fig:normalized-density-plots}
\end{figure}

The results of our investigations are shown in the panels of Fig.~\ref{fig:normalized-density-plots}. 
Each one of them displays the density plots for the Polyakov loop
eigenvalue phases $(\theta_1,\theta_2)$.
Smearing is applied to the configuration before measurements 
to filter the ultraviolet modes that are not relevant for the location of the minima of $V_{\rm eff}$.

Although a bit
noisy because of the procedure, it shows the expected features of the
perturbative potential, and can be directly compared with the perturbative prediction of
Fig.~\ref{fig:ad-mass}.
The plots, from left to right, are respectively the $X$, $A$,
$B$, and $C$ phases. The distribution 
in the $X$ (confined) phase is a constant {\it i.e.} unity so the plot 
is a white image, which is a manifestation of a uniform random
distribution of the eigenvalues in the two dimensional plane.
An interesting finding is that the $C$ phase shows a completely
different behavior from the confined one. The eigenvalues are now not
distributed in a random fashion but located in peaks around the $Z_3$
symmetric values $\theta_i = 0, \pm 2\pi/3$ (again some artifacts
appear), with maximal repulsion between them (see a semi-classical analysis e.g. in~\cite{Poppitz:2012nz}).
All the four predicted phases are clearly represented by the data, which
is a strong indication of the realization of the Hosotani mechanism
in 3+1 dimensions even at the non-perturbative level.

Numerical simulations are performed on the Hitachi SR16K at Kyoto University and the SR16K and the IBM System Blue Gene Solution at KEK
under its Large-Scale Simulation Program (No. T12-09 and 12/13-23). This work was supported in part 
by grants from the Ministry of Education and Science (No.\ 20244028, 23104009, 21244036).
G.~C and J.~N are supported in part by
Strategic Programs for Innovative Research (SPIRE) Field 5.
H.~H is partly supported by NRF Research Grant 2012R1A2A1A01006053 (HH) of 
the Republic of Korea.

\bibliography{references}{}
\bibliographystyle{lattice}

\end{document}